\documentclass[twocolumn,showpacs,preprintnumbers,amsmath,amssymb,prl]{revtex4-1}

\usepackage[active]{srcltx}
\usepackage{graphicx}% Include figure files
\usepackage{dcolumn}% Align table columns on decimal point
\usepackage{bm}% bold math
\usepackage{color}

\newcommand{\Qb}{{\bf{Q}}}

\renewcommand{\b}{{\bf G}}
\newcommand{\I}{{\rm i}}
\newcommand{\ee}{{\rm e}}

\def\gsim{\lower.35em\hbox{$\stackrel{\textstyle>}{\textstyle\sim}$}}
\def\lsim{\lower.35em\hbox{$\stackrel{\textstyle<}{\textstyle\sim}$}}
\begin{document}

\title{Extraordinary absorption of decorated undoped graphene}

\author{T. Stauber and G. G\'omez-Santos}
\affiliation{Departamento de F\'{\i}sica de la Materia Condensada, Instituto Nicol\'as Cabrera and Condensed Matter Physics Center (IFIMAC), Universidad Aut\'onoma de Madrid, E-28049 Madrid, Spain}

\author{F. Javier Garc\'{\i}a de Abajo}
\affiliation{ICFO-Institut de Ciencies Fotoniques, Mediterranean Technology Park, 08860 Castelldefels (Barcelona), Spain, and ICREA-Instituci\'o Catalana de Recerca i Estudis Avan\c{c}ats, Passeig Llu\'{\i}s Companys, 23, 08010 Barcelona, Spain}

\begin{abstract}
We theoretically study absorption by an undoped graphene layer decorated with arrays of small particles. We discuss periodic and random arrays within a common formalism, which predicts a maximum absorption of $50\%$ for suspended graphene in both cases. The limits of weak and strong scatterers are investigated and an unusual dependence on particle-graphene separation is found and explained in terms of the effective number of contributing evanescent diffraction orders of the array. Our results can be important to boost absorption by single layer graphene due to its simple setup with potential applications to light harvesting and photodetection based on energy (F\"orster) rather than charge transfer.
\end{abstract}

\pacs{78.67.Wj, 78.70.En, 42.25.Bs, 78.20.Ci}

\maketitle

%%%%%%%%%%%%%%%%%%%%%%%%%%%%%%%%%%%%%%%%%%%%%%%%%%%%%%%%%%%%%
%  SECTION INTRODUCTION
%%%%%%%%%%%%%%%%%%%%%%%%%%%%%%%%%%%%%%%%%%%%%%%%%%%%%%%%%%%%%

{\it Introduction.-} The optical properties of graphene have recently been the focus of
special attention due to its potential application to nanophotonics and optoelectronics,
mainly due to the strong electrical tunability displayed by this material over a broad
spectral range down to the infrared \cite{Bonaccorso10}. The absorption of a single layer of undoped graphene takes an approximately constant value $\pi\alpha\approx2.3\%$ over a wide spectral range and is solely governed by the fine-structure constant $\alpha$ and not by material constants
\cite{Nair08,Mak08}. For infrared frequencies, this result is readily obtained within the
linear Dirac model from the universal conductivity $\sigma=e^2/4\hbar$, but it also holds for visible light in spite of trigonal warping effects due to a partial cancellation of the enhanced density of states versus the suppressed dipole moment \cite{Stauber08}. Since the intrinsic light-matter coupling
is given by $\alpha$, one would also expect the absorption of 2D patterned or
decorated graphene to be proportional to this constant.

Despite the low absorption of a single carbon layer, light harvesting based on graphene has been investigated ever since resulting in different methods to increase the level of absorption based on periodic nanopatterning \cite{Ju11,Echtermeyer11,Thongrattanasiri12,Nikitin12,Peres13}, retardation effects \cite{Bludov10,GomezSantos12}, or placing the graphene in a resonant cavity \cite{Ferreira12,Furchi12,Pirruccio13}. Additionally, strong absorption
can also be driven by auxiliary photoactive materials, such as colloidal quantum
dots \cite{Konstantatos12} or semiconducting two-dimensional crystals that transfer electrons/holes into the graphene (e.g., transition metal dichalcogenides such as MoS$_2$ or WS$_2$ \cite{Britnell13}).

The above proposals are limited because they either require advanced experimental facilities or operate over a small spectral range. In this Letter, we propose a simple method applicable over a wide spectral range using graphene as photoactive material. The structures under consideration (e.g., randomly depositing non-absorbing scatterers on top of an undoped graphene layer) only involve standard fabrication techniques, and they can be tested using conventional optical characterization setups.
The absorption is mediated by energy transfer into the graphene (F\"orster effect \cite{Forster48}), which exploits the excellent quenching properties of this material \cite{Kim10,Chen10,Loh10} to strongly redirect the evanescent field produced by the small scatterers into the absorbing carbon layer. Also crucial for the efficiency of the proposed mechanism is the algebraic dependence of absorption on distance $z$ as $\sim z^{-4}$ for undoped graphene \cite{Swathi09,GomezSantos11,Gaudreau13}, in contrast to the exponential decay with distance of plasmon-driven absorption near doped graphene \cite{Velizhanin11}. This allows us to consider distances $z$ for optimum absorption of the order of (weak scatterers) or well beyond (resonant scatterers) the Fermi wavelength $\lambda_g=\lambda v_F/c$, where $v_F\approx10^6$m/s is the Fermi velocity and $\lambda$ is the optical wavelength. Alternatively, the increase in absorption associated with the presence of surface scatterers can be qualitatively interpreted as the result of light spending more time near the graphene layer \cite{Atwater10}.

{\it Absorption driven by an individual dipole.-}  We first discuss a single
non-absorbing particle and show that in the presence of graphene the extinction
cross-section is partially converted into an absorption cross-section. The
particle is described through a point dipole excited by an incoming light
field. In the absence of graphene, the particle scatters light with an
extinction cross-section $\sigma^0_{\rm ext} = \frac{k}{\epsilon_0} {\rm
Im}\{\alpha_p\}$, where $k=2\pi/\lambda$ and $\alpha_p$ is the particle
polarizability. For a weak scatterer, we can approximate $\alpha_p^{-1} =
\alpha_E^{-1}-\I k^3/6\pi\epsilon_0$, where $\alpha_E$ is the real
electrostatic polarizability and the $k^3$ term is introduced to satisfy the
optical theorem to first order \cite{Draine94}, so that we have $\sigma^0_{\rm ext} =(\alpha_E^2/6\pi
\epsilon_0^2)k^4$. In the strong scatterer limit (e.g., a perfect two level
atom), the dipole strength is only limited by radiation reaction and we can
approximate $\alpha_p^{-1}\approx-\I k^3/6\pi\epsilon_0$. This yields
$\sigma^0_{\rm ext}=6\pi/k^2\sim\lambda^2$, which is large compared with the
cross section typically exhibited by metallic nanoparticles.

In the presence of  graphene, the decay rate is largely modified and part of the released energy is absorbed
by the carbon layer, mainly through non-radiative coupling. The absorbed power is
given by $\hbar\omega\gamma_{\rm nr}$, where $\gamma_{\rm nr}$ is the  non-radiative decay rate. In units of the natural radiative decay rate $\gamma_0$, and only considering the longitudinal response, which is dominant at short 
distances $z\ll\lambda$, we find a characteristic distance dependence \cite{Swathi09,GomezSantos11,Gaudreau13}
\begin{align}
\tilde{\gamma}_{\rm nr}=\frac{\gamma_{\rm nr}}{\gamma_0}=  \frac{9\alpha}{256\pi^3} \; \frac{1}{(1 +\epsilon)^2} 
\frac{1}{(z/\lambda)^4},
\end{align}
where $\epsilon$ is the dielectric constant of the substrate, assumed to be real. This non-radiative decay saturates for $z\lessapprox\lambda_g$, reaching extremely large values $\tilde{\gamma}_{\rm nr} > 10^6$. 

\begin{figure} % Requires \usepackage{graphicx} 
\includegraphics[clip,width=8cm]{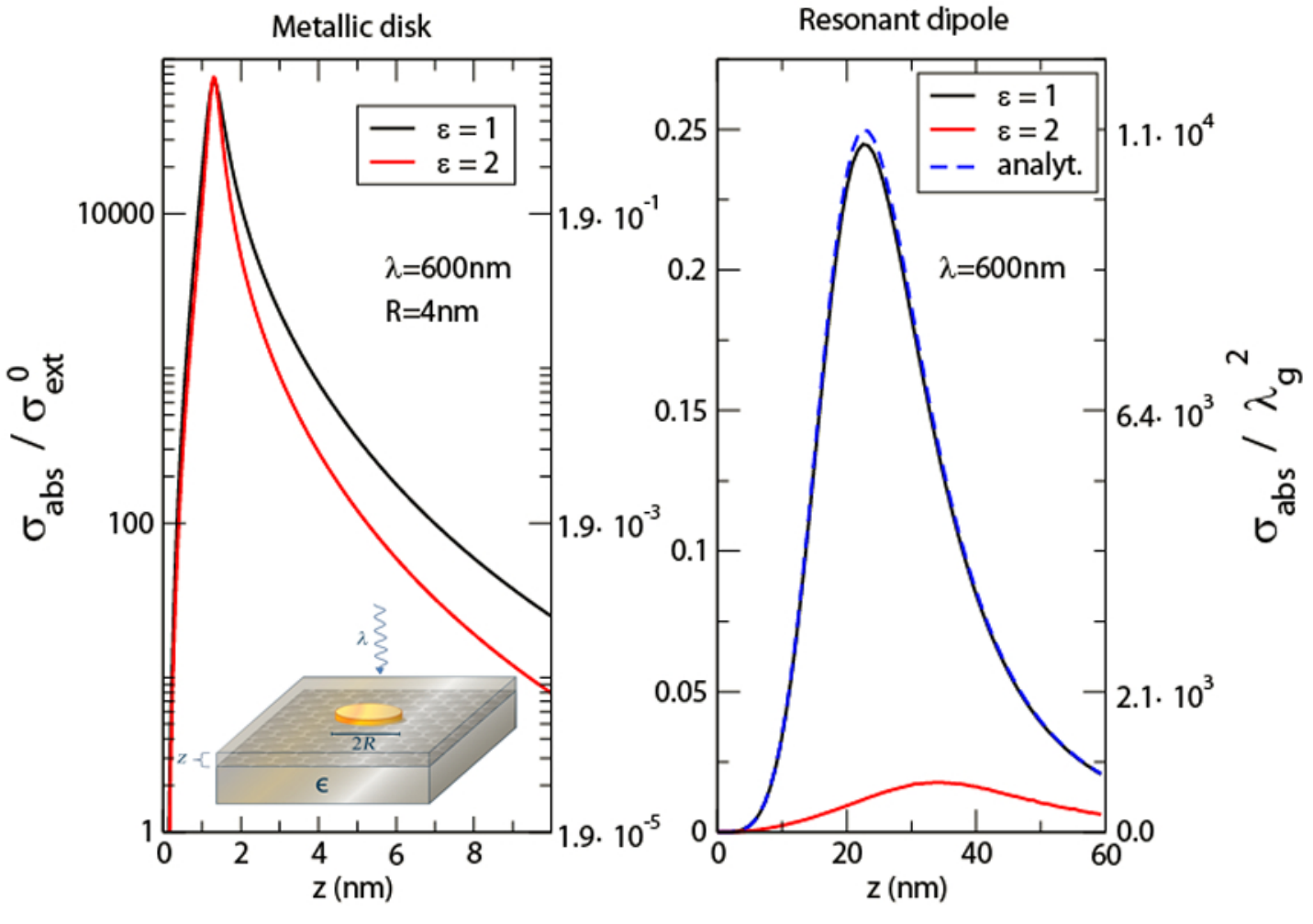} \\
 \caption{(color online) Left panel: dependence of the absorption cross-section by a
non-absorbing metallic disk (radius $R = 4 \,\text{nm}$) on its distance to an undoped
graphene layer, either suspended ($\epsilon=1$) or supported on glass ($\epsilon=2$) for normal incidence. The
light wavelength is $\lambda= 600 \,\text{nm}$. The absorption cross-section is given
relative to both the bare scattered extinction (left vertical scale) and the 
graphene units
(right scale) with $\lambda_g=\lambda v_F/c$ Right panel: same as left panel for a resonant dipole. Full theory (solid curves) is
compared with the analytical approximation of Eq.\ (\ref{analytical}) (dashed curves).}
\label{crossweak}
\end{figure}

If the dipole excitation was decoupled from the graphene, the absorption cross-section could be largely enhanced by simply bringing the dipole closer to
the carbon layer, leading to $\sigma_{\rm abs}\rightarrow\sigma^0_{\rm ext}\tilde{\gamma}_{\rm nr}$. However, the oscillating dipole amplitude is affected by the graphene through strong back-reaction mediated by evanescent waves. The self-consistent electric field induced at the position of the dipole particle becomes ${\bf E}^{\rm ind}=G{\bf p}$, where $G$ is a particle-independent Green function (see Supplementary Information, SI \cite{supplmaterial}). This allows us to write the long-wavelength limit of the normal-incidence absorption cross-section as
\begin{equation}\label{cross} 
\sigma_{\rm abs} \approx \frac{k}{\epsilon_0} \left(\frac{2}{1+\sqrt{\epsilon}}\right)^2 \frac{{\rm Im}\{G\}}{|\alpha_p^{-1} - G|^2},
\end{equation}
which is linear in ${\rm Im}\{G\}\propto {\rm Re}\{\sigma\}$ (i.e., the real part of the graphene conductivity $\sigma$). For a weak perfect-conductor disk scatterer (Fig.\ \ref{crossweak}, left panel), we observe a large absorption exceeding $10^4$ times the bare disk extinction cross-section. The simple expression
$\sigma_{\rm abs}\rightarrow\sigma^0_{\rm ext}\tilde{\gamma}_{\rm nr}$ works well at large distances, as back-reaction effects can be disregarded for weak scatterers. The expected saturation appears at rather small distances, where the dipole model is anyway a poor approximation. For a strong scatterer (Fig.\ \ref{crossweak}, right panel), assuming a perfect dipolar resonance with $\alpha_p^{-1} = -\I k^3/6\pi\epsilon_0$, we obtain a sizable absorption reaching $25\%$ of the bare-scatterer extinction cross-section. A strongly non-monotonic behavior with separation is then observed due to back-reaction, leading to a maximum absorption at much larger distances than in the weak scatterer regime.

As anticipated above, the absorption is given by expressions involving the fine-structure constant for distances $z\gtrsim\lambda_g=\lambda v_F/c $, the regime where the graphene conductivity can be substituted to an excellent approximation by its local, universal limit, $\sigma=e^2/4\hbar$. For example, for a resonant scatterer in front of free standing graphene, we find (see SI \cite{supplmaterial})
\begin{equation}\label{analytical} 
 \frac{\sigma_{\rm abs}}{\sigma^0_{\rm ext}}\approx  \frac{9\alpha}{1024\pi^3}
 \frac{(z/\lambda)^4}{\left[(z/\lambda)^4 + \frac{9\alpha}{1024\pi^3}\right]^2},
\end{equation}
which is in excellent agreement with our numerical results, as illustrated in Fig.\ \ref{crossweak} (right panel).

\begin{figure*}
\includegraphics[height=0.49\columnwidth]{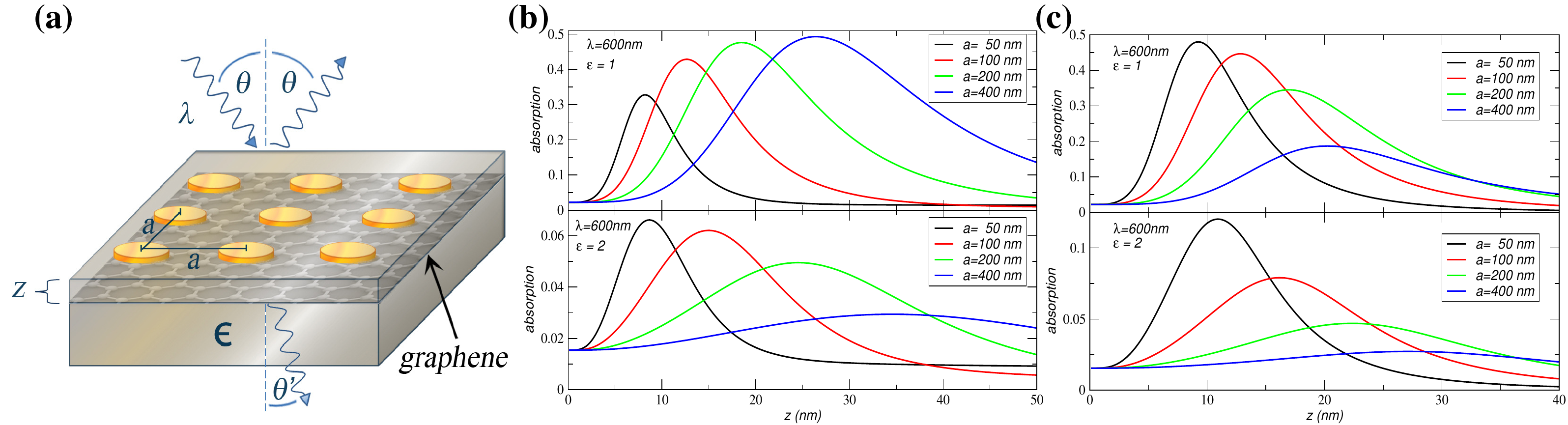}
\caption{(color online) (a) We consider the reflection and transmission of light (wavelength $\lambda$, incident angle $\theta$) in a planar structure consisting of a square array of electric-dipole particles (e.g., metallic disks of radius $R$) placed at a distance $z$ from an undoped graphene layer, which is in turn on top of a substrate of dielectric constant $\epsilon$.
(b) Distance dependence of absorption of light at normal incidence for a
square array of resonant particles and different lattice periods $a$, with  $\lambda= 600\,\text{nm}$ and either supported ($\epsilon=2$, bottom) or self-standing ($\epsilon=1$, top) graphene. (c) Same as (b) for randomly distributed particles with several densities $\rho=a^{-2}$.}
\label{square-resonant}
\end{figure*}

{\it Uniformly decorated graphene.-} 
We consider a layer of uniformly arranged particles that appears to be homogeneous at length scales comparable with the wavelength $\lambda$. In particular, we discuss below both periodic square arrays and completely random distributions, with particle separations $\ll\lambda$. The layer is placed close to a uniform, undoped graphene sheet, and illuminated as indicated in Fig.\ \ref{square-resonant}(a). The parallel wave vector ${\bf k}_{\parallel}$ is fixed by the angle of incidence, leading to a particle dipole amplitude
\begin{equation}
{\bf p}_n = {\bf p}({\bf k}_{\parallel}) \; \ee^{\I {\bf k}_{\parallel} \cdot {\bf r}_n},
\end{equation}
where ${\bf r}_n$ is the position of particle $n$. Then, we can write
\begin{equation}\label{polariz}
{\bf p}({\bf k}_{\parallel}) = \left[{\bm \alpha}_p^{-1} - {\mathfrak G} ({\bf k}_{\parallel})\right]^{-1} {\bf E}^{\rm ext},
\end{equation}
where ${\bf E}^{\rm ext}$ is the external electric field,
\begin{equation}\label{self}
 {\mathfrak G}({\bf k}_{\parallel})\! = \!\! \int\! \frac{d^2\Qb}{(2 \pi)^2}%\\ 
 \left[\frac{\mathcal S(\Qb -{\bf k}_{\parallel} )}{\rho}
 [1 +  {\mathfrak r(\Qb)\,\ee^{2 \I k_z z}}]\, -1\right] {\mathfrak g}(\Qb) \end{equation}
describes the dipole-dipole interaction, $k_z=\sqrt{k^2-Q^2}$, $\rho$ is the particle density, ${\mathfrak r}$ is the diagonal graphene/substrate reflection matrix depending on the polarization of the incident wave, and ${\mathfrak g}$ is the vacuum Green function. The explicit definitions are given in the SI \cite{supplmaterial}. 

All information on particle positions is contained in $\mathcal S$, the Fourier transform of the dipole-pair correlation function (see below). The absorbance can then be calculated from the power absorbed per unit area by graphene (see SI \cite{supplmaterial}). 
 
{\it Periodic particle array.-} 
The structure factor in a periodic array reduces to the contribution from reciprocal lattice vectors $\b$,
\begin{equation}\label{Sarray}
\mathcal S(\Qb) = (2 \pi\rho)^2 \sum_{\b} \delta(\Qb - \b).
\end{equation}
The integral in Eq.\ (\ref{self}) is then transformed into a sum over discrete vectors $\Qb={\bf k}_\parallel+\b$, which we evaluate for square arrays of period $a$ (see \cite{deAbajo07,supplmaterial} for more details).

In the non-diffractive regime (i.e., for $\lambda>a$), the absorbance reduces to
\begin{align}
{\cal A}_j=1-|\tilde r_j|^2-|\tilde t_j|^2\frac{k_z'}{k_z},
\end{align}
where $j=p,s$ indicates the light polarization. These corresponding Fresnel coefficients $r_j$ and $t_j$ are the sum of direct (without particle dipoles) and radiated (super index $d$) contributions, 
$\tilde r_j=r_j^{d}+r_j\;,\;\tilde t_j=t_j^{d}+t_j$, 
which we derive from Eq.\ (\ref{self}) following the methods of 
Ref. \cite{Thongrattanasiri12}. 
Explicit coefficients for these coefficients are given in the SI \cite{supplmaterial}. 

%%%%%%%%%%%%%%%%%RESULTS%%%%%%%%%%%%%%%%%%%%%%%%%%%%%%%%%%%%%%%

For weak scatterers, the absorption is small and quickly vanishing with the distance
between the array and the graphene (see SI \cite{supplmaterial}). We thus concentrate
on the more interesting case of resonant dipoles. Results for several lattice densities
close to either suspended or silica-supported graphene are shown in Fig.\ \ref{square-resonant}(b). One observes a large enhancement in
absorption, nearing $\sim50\%$ for free-standing graphene. Again, the non-monotonic behavior with
distance originates in the screening effect of graphene reflection at small
separations. Particularly striking is the behavior with particle density $\rho$ for
free-standing graphene [Fig.\ \ref{square-resonant}(b), top]: the overall absorption increases with decreasing
density over a significant range of distances where absorption is high. 
As the absorption coefficient ${\cal A}$ is dominated by evanescent modes, the
following
expression can be obtained in analogy with Eq.\ (\ref{cross}) under normal incidence:
\begin{equation}\label{rhocross}  
{\cal A} \sim {\cal A}_{\rm ev}  \approx \rho
\frac{k}{\epsilon_0} \left(\frac{2}{1+\sqrt{\epsilon}}\right)^2 \frac{{\rm Im}\{\Delta
G_{\rm ev}\}}{|\alpha_p^{-1} - G|^2}, 
\end{equation} 
where $\Delta G_{\rm ev} $ is the evanescent-wave contribution to Eq.\ (\ref{self}) with
$|\Qb| > k$. 
From Eq.\ (\ref{rhocross}), one would naively expect a linear monotonic dependence with density
at fixed distance. Notice however that $\Delta G_{\rm ev} $ also depends on density. In
fact, Eq.\ (\ref{Sarray}) amounts to a discrete sampling of the integral in Eq.\ 
(\ref{self}), from which the specular contribution is removed. The exponential term
$\ee^{\I 2  k_z z}$ ($=\ee^{-2  |k_z| z} \approx\ee^{-2 |\b| z} $ for $\b\neq0$) in Eq.\ (\ref{self}) effectively limits the sum to $|\b| \lesssim
1/z $, whereas the lowest (evanescent) contributing term is $|\b_1|= 2\pi/a $.
Therefore, a reduction in particle density (and hence, also in $|\b_1|$) produces an
increase in the number of evanescent modes effectively contributing to absorption.

{\it Disordered layer of dipoles.-}
In a totally disordered array, the structure factor becomes
\begin{equation}\label{random}
\mathcal S(\Qb) = (2 \pi\rho)^2  \delta(\Qb) + \rho.
\end{equation}
The dipolar response now reduces to the diagonal elements
\begin{equation}\label{selfrandom}
(\alpha_p^{-1} - G)_{\parallel,\perp} = 
\alpha_{p}^{-1} 
- {G}^0_{\parallel,\perp}
-\Delta{G}_{\parallel,\perp},
\end{equation}
where $G^0$ is the contribution from the $\b=0$ specular term and $\Delta G$ originates in the continuum of evanescent modes, as given by
\begin{equation}%\label{DeltaGrandom}
 \Delta {G}_{\parallel} = \Delta{G}_{\perp}/2= 
 \frac{-1}{(2\pi)^2} \int d^2\Qb \,r_p^{\rm nf}(\Qb)\, {\mathfrak g}_{xx}(\Qb) \ee^{-2 Q z}.
\end{equation}
Here, we use a non-retarded approximation for the reflection coefficient $r_p^{\rm nf}(\Qb) $, consistent with the short-distances $z\ll\lambda$ considered.

\begin{figure} % Requires \usepackage{graphicx} 
\includegraphics[clip,width=8cm]{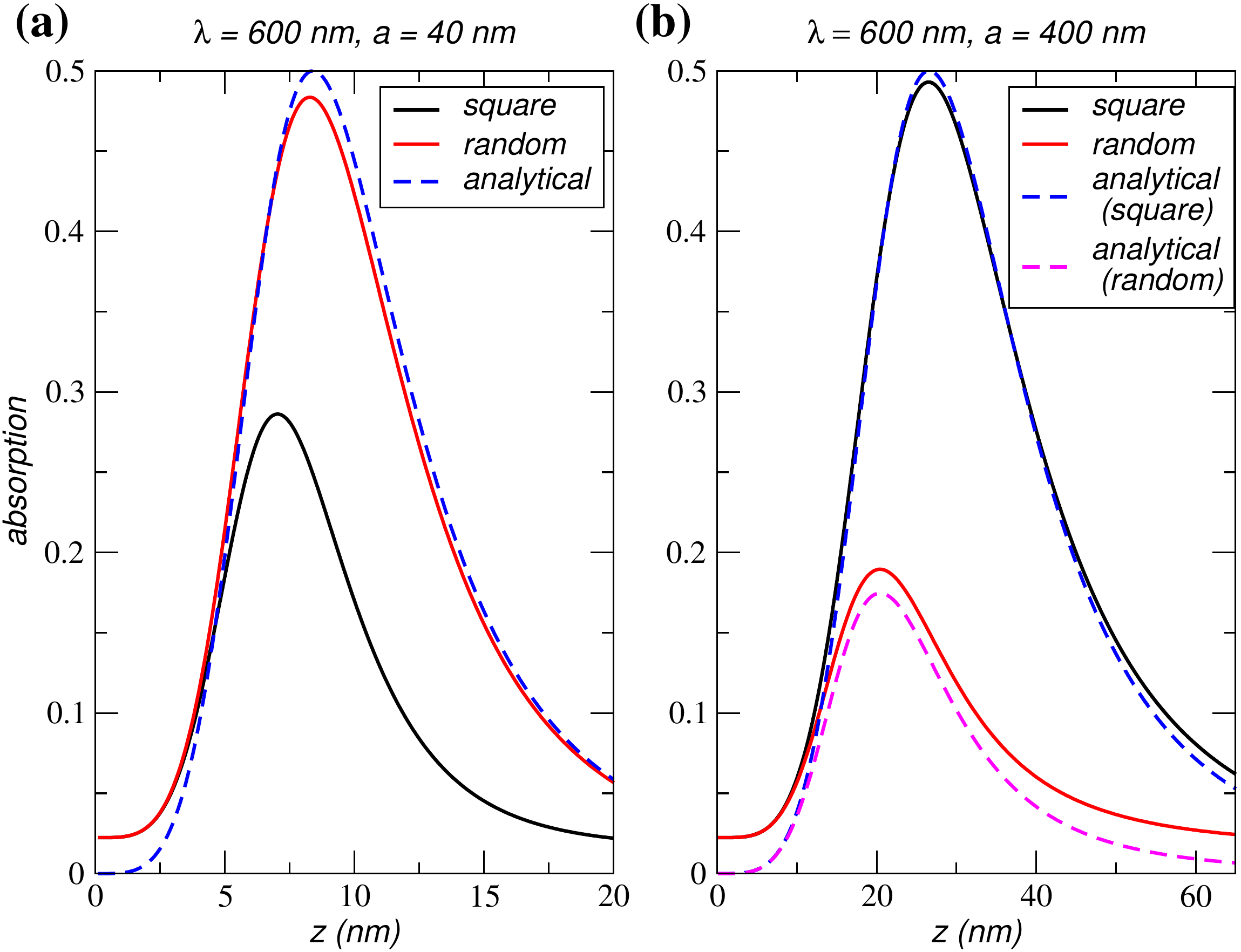} \\ 
\caption{(color online) Left panel: Dependence of absorption on the distance $z$ between either an ordered or a disordered particle array with respect to the graphene [see Fig. \ref{square-resonant}(a)]. The particle density is
$\rho = 40^{-2}\text{nm}^{-2}$ in both cases, the graphene is self-standing and the light wavelength is $\lambda= 600 \,\text{nm}$. Full results are compared with the analytical expression of Eq.\ (\ref{analytical2}). Right panel: same as left panel for a density $\rho = 400^{-2}\text{nm}^{-2}$.}
\label{resonant-comparison}
\end{figure}

As expected, the absorption for weak scatterers (not shown) is very similar in ordered and disordered lattices. For the resonant case, the absorption is shown in Fig.\ \ref{square-resonant}(c). Although broadly similar, several differences are noticeable between ordered and random arrays, specially in the low and high particle-density limits. This comparison is highlighted in Fig.\
\ref{resonant-comparison} for free-standing graphene, showing that random arrays produce more absorption than ordered ones at high densities, reaching almost $50\%$. The opposite happens at low densities, where ordered arrays produce higher absorption, again around $50\%$.

We compare the absorbance at resonance frequency of both ordered and disordered free-standing graphene with the analytical approximation 
\begin{equation}\label{analytical2} 
 {\cal A} \approx  
 \frac{2 B (z/\lambda)^4}{[C(z/\lambda)^4 + B]^2},
\end{equation}
with $B=3\alpha/(256\pi^2\rho\lambda^2)$, and either $C = 1$ (ordered array) or  $C = 1+4\pi/(3\rho\lambda^2)$ (disordered array). Equation\ (\ref{analytical2}) is obtained from Eq.\ (\ref{selfrandom}) by setting  ${\rm Re}\{\alpha_p^{-1}\}=0 $ (resonant case), using the local universal value for the graphene conductivity $\sigma=e^2/4\hbar$, valid for distances $z\gtrsim\lambda_g=\lambda v_F/c $. Interestingly, only $\alpha$, the scaled distance $z/\lambda$, and the number of particles per square wavelength $\rho\lambda^2$ appear in Eq.\ (\ref{analytical2}), i.e., no material constants of graphene enter and the proposed mechanism is independent of the dipole resonance $\Omega=2\pi c/\lambda$ due to graphene's broadband properties. The line width of the isolated dipole resonance $\gamma\ll\Omega$ can further be
strongly enhanced due to absorption in the graphene layer to
$\gamma_{abs}=\frac{3}{2\pi}\rho\lambda\gamma$, yielding extremely large values for
high concentrations and long resonance wavelengths, see SI.

Fig. \ref{resonant-comparison} shows the analytical result to be an excellent approximation for random distributions at high densities, whereas  the
numerically calculated absorption of  ordered arrays is substantially lower  due to the detrimental role of the discrete wave-vector distribution of evanescent modes in this limit. In contrast, it is the higher absorption of ordered arrays that almost perfectly matches the corresponding analytical result at low densities.  This can be understood as follows:  the actual particle resonance condition ${\rm Re}\{\alpha_p^{-1}\}=0 $ implies
$\alpha_p^{-1} = -\I k^3/6\pi\epsilon_0$ in Eq.\ (\ref{selfrandom}) for the
disordered array, whereas such an imaginary contribution is absent for the ordered array  due to the layer dynamical self-screening of radiation reaction \cite{deAbajo07}.  In more physical terms, diffuse scattering into radiative modes, absent in (non-diffractive) ordered arrays,  persists in disordered arrays with relative importance increasing at low densities, thus producing stronger radiative response without significantly contributing to absorption. 

We remark that the maximum absorption attainable according to Eq.\ (\ref{analytical2}) is $50\%$, a value closely approached by the numerical results in the regimes discussed above. This is the absolute absorption limit  of a thin layer in vacuum \cite{deAbajo07}, which applies to our structures because $z\ll\lambda$. The dipole arrangements considered, periodic and totally disordered, represent opposite extremes. Therefore, we expect similar absorption enhancements for intermediate situations, such as partially periodic structures, with some degree of disordered, and also for quasicrystalline arrangements.

{\it Summary.-} We have shown that the absorption of a single, undoped graphene layer can be dramatically enhanced (up to $\sim50\%$ in the self-standing configuration) by decorating it with non-absorbing small particles. Both ordered and disordered particle arrays can produce such effect. In a plausible experimental realization, one can deposit small dielectric particles on a graphene layer. Such high absorption is possible over a wide spectral range from the visible to the infrared. A counterintuitive increase in absorption with 
decreasing particle density is predicted for ordered arrays, particularly when resonant particles are considered, which we understand in terms of the effective number of contributing evanescently diffracted orders. An analytical expression in terms of fundamental units has been derived for resonant dipoles, exhibiting a maximum of $50\%$, which is the intrinsic limit for the absorption of a thin layer. To the best of our knowledge, this is the first system to display such behavior for normal incidence without explicitly relying on plasmonic effects \cite{FiftyPercent}. 

Our analysis can be straightforwardly applied to other quasi-2D materials by simply using an appropriate expression for the conductivity $\sigma$, i.e., by replacing $\alpha$ by $\sigma/\pi\epsilon_0c$ throughout our expressions \cite{FewLayer}. The universality of the broadband constancy of the conductivity outside gap regions\cite{Fang13} then yields equivalently universal absorption results in decorated planar absorbing layers and semiconductor heterostructures and multilayer graphene can thus be directly analyzed in this way. Finally, our work provides an alternative strategy to induce photocurrents mediated by energy rather than charge transfer in graphene-based heterostructures, relying on the large optical quenching produced by undoped graphene on nearby optical emitters.

{\it Acknowledgments.} This work has been supported in part by FCT (PTDC/FIS/101434/2008, PTDC/FIS/113199/2009), MIC (FIS2010-21883-C02-02, FIS2012-37549-C05-03), and the EC (Graphene Flagship CNECT-ICT-604391).

\bibliography{fluorabsorb}

\begin{thebibliography}{31}
\bibitem [{Few()}]{FewLayer}%
  \BibitemOpen
  \href@noop {} {}\bibinfo {note} {For (few) $n$-layer graphene, $\alpha$ is
  replaced by $n\alpha$, valid in the optical regime.}\BibitemShut {Stop}%
\end{thebibliography}

\end{document}